\documentclass[aps,prapplied,reprint,superscriptaddress,nofootinbib,longbibliography]{revtex4-2}
\usepackage{graphicx}   
\usepackage{amsmath}
\usepackage{multirow}
\usepackage{booktabs}
\usepackage{xcolor}
\usepackage{csquotes}
\usepackage{comment}

\includecomment{private}

\newcommand{\cevens}{\mbox{CEvNS}}
\newcommand{\skipper}{\mbox{Skipper-CCD}}
\newcommand{\siseroamp}{\mbox{SiSeRO}}
\newcommand{\noise}{\,$\rm e^-_{rms}/pix$}

\begin{document}

\title{CMOS Image Sensors for \cevens\ Detection at Nuclear Reactors}

\author{Santiago Perez}
\email{perezse@uchicago.edu}
\affiliation{Department of Astronomy and Astrophysics, University of Chicago, Chicago, IL, USA}
\affiliation{Universidad de Buenos Aires, Facultad de Ciencias Exactas y Naturales, Departamento de Física, Buenos Aires, Argentina}
\affiliation{CONICET, Universidad de Buenos Aires, Instituto de Física de Buenos Aires (IFIBA), Buenos Aires, Argentina}

\author{Dario Rodrigues}
\email{rodriguesfm@df.uba.ar}
\affiliation{Universidad de Buenos Aires, Facultad de Ciencias Exactas y Naturales, Departamento de Física, Buenos Aires, Argentina}
\affiliation{CONICET, Universidad de Buenos Aires, Instituto de Física de Buenos Aires (IFIBA), Buenos Aires, Argentina}

\author{Miguel Sofo-Haro}
\email{miguelsofoharo@mi.unc.edu.ar}
\affiliation{CONICET, Instituto de Física Enrique Gaviola (IFEG), Comisión Nacional de Energía Atómica (CNEA) and Universidad Nacional de Córdoba, Córdoba, Argentina}
\date{\today}

\begin{abstract}

In this work we present Complementary Metal-Oxide-Semiconductor (CMOS) Image Sensors (CIS) for reactor neutrino experiments targeting coherent elastic neutrino-nucleus scattering (\cevens). We have developed an active-shielding strategy for physical background reduction along with an analytical model describing the impact of intrinsic detector backgrounds from dark current and readout noise at a given frame rate (fps). Assuming a conservative background of 100\,DRU, and a feasible CIS-based detector with $\sim$100\,g of sensitive silicon mass, \mbox{1\,$e^-$} of readout noise, and 200\,fps, it can detect \cevens\ in less than \mbox{50 days} at \mbox{16 meters} from a \mbox{3.6\,GW} thermal reactor core. Improving the detector performance to \mbox{0.2\,$e^-$} readout noise and 1000\,fps substantially reduces the detection time to about \mbox{7 days}. These results establish CIS technology as a promising platform for reactor monitoring and precision neutrino measurements, with thick CMOS sensors featuring parallel and non-destructive readout emerging as particularly attractive candidates for next-generation \cevens\ experiments.
\end{abstract}

\maketitle

\section{Introduction}

Although neutrinos interact only weakly with matter, making their detection inherently challenging, the enhancement of Coherent Elastic Neutrino–Nucleus Scattering (\cevens) cross section at low momentum transfer provides a promising alternative to conventional detection channels. In this process, predicted by the Standard Model in 1974, the neutrino interacts coherently with the entire nucleus via $Z$-boson exchange, resulting in a cross section that scales approximately with the square of the number of neutrons. This enhancement enables neutrino detection with target masses on the kilogram scale and has made \cevens\ particularly attractive for the development of compact detectors for applications such as reactor safeguard and monitoring~\cite{cogswell2016detection, huberfusion, houston2024neutrino, bernstein2020colloquium}. Furthermore, the low nuclear recoil energies involved in reactor \cevens\ measurements, together with the precisely predicted Standard Model cross section, make this process a probe of physics beyond the Standard Model (BSM). In particular, nuclear reactor experiments are sensitive to a variety of new physics scenarios, including light mediators, neutrino magnetic moment measurement, and millicharged particles~\cite{VIOLETA2021, VIOLETA2022, mCP_CONNIE_Atucha}. 

Recently \cevens\ was observed for the first time at a nuclear reactor by the \mbox{CONUS+} experiment using \mbox{3.75\,kg} of germanium detectors~\cite{CONUS+2025}. This milestone was made possible, to a large extent, by the implementation of active shielding systems that veto events coincident with the passage of cosmic muons, thereby suppressing the contribution of muon-induced backgrounds. In contrast, experiments based on silicon detectors employing \skipper\ technology achieve an ultra-low ionization energy threshold ($\sim$5\,eV), substantially reducing the detector exposure (mass or time) required to reach a given sensitivity~\cite{Tiffenberg2017}. Recent results have demonstrated the potential of \skipper\ detectors for neutrino physics~\cite{depaoli2024deployment, 2022_connie} as well as searches for physics beyond the Standard Model (BSM)~\cite{mCP_CONNIE_Atucha}. However, a key limitation of \skipper\ technology is its lack of temporal resolution. Acquiring a single image with a \skipper\ sensor requires a readout time of the order of hours, precluding the use of active shielding systems based on time-coincident vetoes.

In this work Complementary Metal-Oxide-Semiconductor (CMOS) Image Sensors (CIS) are presented as sensors for nuclear reactor neutrino experiment aimed at the observation of \cevens. They have the potential to integrate the advantage of temporal resolution and low ionization threshold in a compact detector. Unlike CCDs, whose serial pixel readout results in long readout times, CIS architectures enable parallel readout of the pixel array, supporting frame rates of up to 1000 frames per second (fps). Furthermore, advances driven by the microelectronics industry have reduced the readout noise of CMOS sensors to below one electron. More recently, \enquote{skipper}-style readout has been demonstrated in CMOS processes, enabling repeated non-destructive measurements and, consequently, an arbitrary reduction of the readout noise.

This work is organized as follows. In Section~\ref{sec:cis}, we review CIS technology, with particular emphasis on recent developments relevant to a CIS-based \cevens\ detector. We then present a strategy for implementing active shielding in such a detector and discuss the expected physical backgrounds. In addition, we introduce an analytical model to estimate the detector's intrinsic background arising from dark current and readout noise. In Section~\ref{sec:detection}, we present the expected \cevens\ signal in silicon targets and evaluate the achievable detection significance for representative nuclear reactor scenarios. In Section~\ref{sec:viable} the expected background and signal models are integrated to assess the performance of a feasible detector design. Finally, Section~\ref{sec:conclusion} summarizes our main findings and discusses their implications for future reactor neutrino experiments. 
\section{CMOS Image Sensors}\label{sec:cis}

\subsection{The CIS Technology}

The Skipper-CCD is a unique image sensor that can achieve deep sub-electron readout noise. The floating gate in the output amplifier enables non-destructive charge measurements, allowing the charge of each pixel to be sampled multiple times. By averaging these repeated measurements, a readout noise as low as \mbox{0.068\noise} can be achieved, approaching the fundamental ionization limit of silicon of \mbox{1.1\,eV} in energy threshold. This capability enables single-electron and single-photon sensitivity at the pixel level~\cite{Tiffenberg2017}. 
In addition, Skipper-CCDs are large-area detectors containing more than one million pixels of \mbox{15\,$\mu$m $\times$ 15\,$\mu$m}. The sensors have a fully-depleted thickness of \mbox{700\,$\mu$m}, corresponding to a sensitive silicon mass exceeding \mbox{0.5\,g} per device. The combination of a very low ionization threshold and a large target mass in Skipper-CCDs enables rare-event physics experiments. Thus, those employing \skipper\ detectors have achieved world-leading sensitivity across a broad range of \mbox{sub-GeV} dark matter masses~\cite{PhysRevLett.121.061803,PhysRevLett.122.161801,PhysRevLett.125.171802,PhysRevLett.130.171003}. Moreover, they are being used to probe new physics in reactor neutrino experiments~\cite{depaoli2024deployment,2022_connie,mCP_CONNIE_Atucha,aguilar2024searches}.

In conventional CCDs, all pixels are read out sequentially through a single amplifier~\cite{janesick}, which inherently limits the readout speed. In \skipper\, the readout time is further increased by the need to acquire multiple samples per pixel. The requirements of future instruments, such as the Stage-5 Spectroscopic Survey Experiment and the Habitable Worlds Observatory, have therefore motivated efforts to increase the readout speed of \skipper\ sensors~\cite{schlegel2022spectroscopic,tech_gap,rauscher2022radiation}. Two complementary output-stage strategies have been proposed to address this limitation. The first is the multiple-amplifier sensing CCD (MAS-CCD), which integrates more than 16 floating-gate amplifiers per CCD, increasing the readout speed proportionally~\cite{botti2024single}. The second is the Single-Electron Sensitivity Readout (\siseroamp) amplifier, which replaces the floating gate with an integrated MOSFET featuring an internal gate. This design improves at least six times the readout speed of the floating-gate amplifier~\cite{sofo2024achieving}. In addition, dual-sided charge-coupled devices have been proposed to improve background rejection capabilities~\cite{tiffenberg2024dual}. Nevertheless, even with improved output stages, the inherently serial architecture of the CCD pixel array remains a fundamental limitation for further increases in readout speed. One of the main consequences of this limitation is the impossibility of implementing active shielding for background reduction, which is crucial for rare-event physics experiments, particularly those operating in surface laboratories.

On the other hand, CMOS image sensors (CIS) have come to dominate the consumer imaging market, benefiting from widespread fabrication infrastructure and advanced processing options. In a CIS, each pixel includes a sensing element—typically a photodiode—coupled to an output amplifier, usually a \mbox{MOSFET} operating as a source follower. The sensing element collects the charge during the exposure period, followed by a charge transfer operation to the output amplifier~\cite{cmos_fossum}. The inherently parallel readout of pixels allows for higher frame rates than CCDs. The scaling of CMOS technology to feature sizes below 100\,nm has enabled pixels with extremely low capacitance, resulting in high sensitivity and low readout noise. Noise values below \mbox{1\,$\rm e^-$} for CMOS image sensors are regularly reported~\cite{seo20150, boukhayma2016sub}. One case that exploits the miniaturization of CMOS process and exhibits \mbox{photon-counting} capability is the Quanta Image Sensor (QIS)~\cite{Quanta_Fossum,QIS_whitepaper}. The QIS architecture is based on binary pixels known as \enquote{jots}. These sensors achieve 0.19\,$\rm e^-$ readout noise at 30\,fps with 16\,Mpix frames and a pixel size of 1.1\,$\mu$m. In 2021, Hamamatsu introduced a commercial camera based on its \enquote{quantitative CMOS} (\mbox{qCMOS}) technology. This sensor achieves a readout noise of 0.27\,$\rm e^-$ at 5\,fps and 0.43\,$\rm e^-$ at 120\,fps, for a 9.6\,Mpix array with 4.6\,$\mu$m pixels~\cite{hamamatsu_technote}. 

In addition to low readout noise, rare-event physics experiments also require a substantial target silicon mass. High-voltage CMOS technologies originally developed for automotive applications have been used to implement depleted monolithic active pixel sensors (DMAPS) using high-resistivity, thick substrates. These technologies have been primarily developed for the inner tracking detectors of high-energy colliders, as they allow of large area detectors for relatively low price~\cite{snoeys2014cmos}. Several CMOS processes have been explored for this purpose, including the 180\,nm TowerJazz CMOS image sensor process~\cite{rinella2017alpide,senyukov2013charged}, the \mbox{150\,nm} and \mbox{110\,nm} LFoundry process~\cite{hirono2016cmos,wang2017development,Pantouvakis_2026,ARCADIA}, and the \mbox{AMS 180\,nm} technology~\cite{peric2012active}. DMAPS \mbox{100\,$\mu m$} thick, fabricated in \mbox{2\,k$\Omega \cdot$ cm} substrate, can be fully depleted at reasonable voltages of 20\,V\cite{hirono2016cmos}. In particular, the ALPIDE chip is fabricated in high resistivity epitaxial layer (\mbox{$>$1\,k$\Omega \cdot$ cm}) in the range of 18 to 30\,$\mu$m thick~\cite{acharya2024alice}. Each chip measures $\rm 15\times30\,mm^2$ with integrated electronics that allows to handle Pb-Pb collisions at 50$\,$kHz. The ALICE inner tracker system is composed of 24000 ALPIDE chips and leverages the high-volume manufacturing capabilities of commercial CMOS foundries~\cite{triolo2023calibration}. Another application field where DMAPS are developed is for X-ray free-electron laser (XFEL), where soft X-ray detection is required at very high frame rates. One example is the \mbox{FLORA} detector developed for the \mbox{LCLS-II}~\cite{carini2019hybridized}. This sensor can reach frame rates of up to 10000 frames per second and is sensitive to photons with energies as low as \mbox{250\,eV}, with a readout noise of approximately \mbox{10\,$e^-$}~\cite{carini2019hybridized}. An interesting example is the \mbox{CMOS-DEPFET} developed for the European XFEL~\cite{cmos-depfet}. The sensitive device, composed by an array of \mbox{128\,x\,512} pixels, was fabricated in a high-voltage \mbox{350\,nm} CMOS process~\cite{cmos-depfet-castoldi}\cite{cmos-depfet-first}. The device has been demonstrated to operate at \mbox{1.1\,MHz} frame rate with \mbox{9.8\,$e^-$} in a \mbox{725\,$\mu m$} fully depleted thick substrate at $18\,^\circ\mathrm{C}$\cite{cmos-depfet}\cite{cmos-depfet-porro}. The pixel array is bump-bonded to a readout ASIC fabricated in a \mbox{130\,nm} CMOS technology~\cite{cmos-depfet-integration}. Each pixel is a depleted MOSFET featuring an internal gate for charge measurement called DEPFET~\cite{depfet}. DEPFETs allows non-destructive charge readout and has been demonstrated to reach \mbox{0.2\,$e^-$} in a \mbox{64\,x\,64} array developed for dark matter searches~\cite{depfet-darkmatter}. The \siseroamp\ amplifier is based on the same amplification principle as the DEPFET and has been demonstrated for the readout of fully depleted thick CCDs, achieving a noise performance of \mbox{0.25\,$e^-$} with pixel readout times below 1\,ms~\cite{sisero}. Therefore parallel readout an array of pixel at \mbox{1000\,fps} with sub-electron noise is feasible.

Another recent development is the \mbox{Skipper-CMOS} architecture, which demonstrates \mbox{Skipper-style} non-destructive charge readout implemented with a floating-gate amplifier in Tower Semiconductor's $180\,\mathrm{nm}$ CMOS image sensor process~\cite{lapi-skipper-cmos}. The pixel consists of a pinned photodiode coupled to a floating-gate amplifier through a transfer gate. Sub-electron readout noise is achieved after averaging twenty samples acquired within $10\,\mu s$. Because the measurement is non-destructive, the noise can be reduced arbitrarily by increasing the number of samples, reaching $0.135\,e^-$ when averaging one thousand samples. This resolution is maintained across the entire full well of the device, which is approximately $1.5\times10^{4}\,e^-$. This feature is key for applications in which ionization processes, such as \cevens, produce multiple \mbox{electron-hole} pairs on timescales that are effectively instantaneous compared with the readout time. The design is compatible with backside illumination using $18\,\mu\mathrm{m}$-thick epitaxial wafers with a resistivity of $1000\,\Omega\cdot\mathrm{cm}$~\cite{stefanov2016fully}. Although the dark current of this specific device has not yet been reported, values on the order of $0.15\,e^-$/pixel/s at $-20^\circ$C are expected based on comparable CMOS image sensors fabricated in similar processes~\cite{ma20154mp,wang20154m}. To enable high frame-rate operation, a dedicated readout integrated circuit (ROIC), SPROCKET, has been developed~\cite{quinn2024cryogenic}\cite{quinn2023cryogenic}. SPROCKET is implemented in a $65\,\mathrm{nm}$ CMOS process and is designed to interface with the Skipper-CMOS pixel matrix through hybrid integration using micro-bump interconnects.

In this work, we propose CIS for neutrino detection. We consider that the technology is sufficiently mature to enable the fabrication of thick sensors, with low readout noise, high frame rates in commercial CMOS foundries. As presented below, high frame rates provide the opportunity to implement active shielding techniques for background suppression. In the following section, we derive the requirements on the threshold, the sensitive silicon mass, and the time resolution that a CIS must satisfy to enable the observation of \cevens\ in nuclear reactor experiments.

\subsection{Active Shielding in a CIS-Based Detector}

Atmospheric muons are the products of the interaction of cosmic rays with the atmosphere. These muons can interact with detector material and produce neutrons that, after moderation, can interact with the detector adding background in the signal range. Dark matter searches experiments are required to be deployed in underground laboratories in order to fundamentally reduce this background, but in reactor neutrinos detection it is not a possibility and therefore passive and active background reduction techniques have to be implemented. The active shield technique is applied in detectors that have a pulse response every time that a particle interaction carries on. When there is a time coincidence between the pulses of the muon detector and the signal detector, the event is vetoed. The time coincidence is set in a temporal window on the order of tens of microseconds estimated from the moderation time of the neutrons. This technique can significantly suppress the background induced by muons.

One experiment that has been taken as a reference case for this work is CONUS+, being the first experiment that has detected \cevens\ in a nuclear reactor~\cite{CONUS+2025}. The experiment consist of four high purity germanium detectors with \mbox{3.75\,kg} of active mass and an energy threshold of \mbox{210\,eVee}. It is located at the Leibstadt Nuclear Power Plant in Switzerland (KKL), inside the reactor building that provides an average overburden of \mbox{7.4\,m} of water equivalent \mbox{(m w.e.)}\cite{ackermann2024conus+}. The detector is placed \mbox{20.7\,m} from the reactor core. The whole shield has a total volume of \mbox{$\rm 1.65\,m^3$}, with a total of \mbox{20\,cm} of lead in all directions. The background is primarily dominated by prompt muon-induced events\cite{CONUS+2025}. The active shield consist of two internal layers of scintillation plates installed as an active muon veto. 
The active shield reduce the background four orders of magnitude reaching \mbox{$\rm\sim\,10\,counts\,kg^{-1}\,d^{-1}$} in the \mbox{$[0.4,1]\,$keV$_{ee}$} energy region, being fundamental to enable the detection of \cevens~\cite{bonet2023full}. 
The total muon rate measured by the full inner veto at KKL was \mbox{($\rm274 \pm 1)$\,Hz}~\cite{ackermann2024conus+}. 

It is important to mention that shield volume is mainly constrained by the volume of the inner germanium detectors. A reduction on the volume of the sensitive detector would imply a smaller shielding and therefore a lower \mbox{$\mu$-rate}.

CIS do not respond instantaneously to radiation; instead, they integrate the charge produced by ionization over a finite exposure time before readout, which defines the frame rate. CIS enable parallel pixel readout, allowing frame rates of hundreds to thousands of frames per second. While this readout scheme does not allow a direct implementation of the muon veto described above, we instead propose to exploit the counting Poisson statistics of cosmic muons together with a sufficiently high frame rate. Frames acquired in coincidence with a trigger from the muon detector are rejected. As we show below, this approach allows a live time exceeding 90\%.

The fraction of images that are rejected when there was a trigger from the muon detector determines the effective live time of the experiment and is closely related to the frame rate. Assuming that muon arrivals follow a Poisson process, the probability that no muon crosses the detector during the acquisition of a single image is

\begin{equation}
P(\text{zero muons per image}) = e^{-\lambda}
\label{eq:poisson_muons}
\end{equation}
where $\lambda$ is the expected number of muons crossing the vetoed volume during the exposure time of one frame. 
Equation 1 assumes single frame rejection, in which only the frame coincident with the muon trigger is discarded. Since the moderation and capture of muon-induced neutrons proceed over a finite time $\tau_n$ $\sim$ 450~$\mu$s as reported by CONUS, a muon arriving near a frame boundary can produce neutron activity that extends into the following frame. A more conservative veto could reject additional consecutive frames, at the cost of a reduction in live time. At the frame rates considered here, where the frame time is large when compared with the moderation time, this effect is small and single-frame rejection provides a reasonable approximation.

For a CIS in a shielding configuration similar to the one of the \mbox{CONUS+} experiment, the total muon rate measured by the full inner veto was $(274\pm1)\,\text{Hz}$~\cite{ackermann2024conus+}. Since the muon crossing rate scales with the transverse area of the vetoed volume, the smaller sensitive volume of the proposed CIS-based detector translates into a proportionally lower rate, and we adopt a representative value of $\sim$200\,Hz for the following estimates. Under these assumptions, the expected number of detected muons per frame acquisition time is:
\[
\lambda = \frac{200\,\text{Hz}}{fps},
\]
\noindent where $fps$ is the frame per second acquisition rate. Therefore with Eq.~\ref{eq:poisson_muons} it is possible to calculate the probability of accept a frame. This probability can be directly related to detector live time. For example, at 1800\,$fps$, a 90\% of the frames are accepted, the acquisition time of the accepted frames is the detector live time. It is important to mention that a regime similar to the \mbox{CONUS+} experiment is achieved in this condition. In \mbox{CONUS+} active shield, a rejection window of 450\,$\mu s$ (equivalent to 2222 $fps$) is open every muon trigger resulting in $\sim12\%$ of dead time ~\cite{ackermann2024conus+}.
In the case of CONUS experiment at Kernkraftwerk Brokdorf after applying the muon veto the background is reduced four orders of magnitude down to \mbox{16.6\,DRU} in the $\rm [0.4,1]~keV_{ee}$ energy range dominated by environmental gamma rays ~\cite{ackermann2024conus+}. At low deposited energies, gamma interactions in silicon proceed primarily through Compton scattering, producing electron recoils with a slowly varying differential spectrum. Therefore below 100\,eV, that is the energy range relevant for \cevens, the background can be approximated as flat, as supported by dedicated Compton and background measurements with \skipper~\cite{botti2022constraints, MoroniFernandez:2022eie}. Therefore, motivated by this expectation, in the following calculations a \textit{flat background of \mbox{100\,DRU} will be assumed} for the case of a CIS-based \cevens\ detector. It is worth noting that eliminating the neutron contribution through the active shield suppresses the emergence of an exponential low-energy physical background.

Cherenkov photons are emitted during the passage of cosmic-ray muons through silicon~\cite{Du:2022prx}. These photons can subsequently ionize the silicon, generating an additional source of background within the signal energy range~\cite{guille_muones}. To mitigate this effect, dark matter and neutrino detection experiments employing CCDs typically apply a fiducial cut, excluding a region of pixels surrounding each muon event~\cite{PhysRevLett.122.161801}. The primary drawback of this approach is the associated loss of active detector mass. As the CCD readout time increases, the number of muons per frame also rises, leading to a larger fraction of the detector being discarded and, consequently, a greater effective mass loss. In addition to the application of an active shield, increasing the frame rate provides a further advantage: frames containing muon events can be more efficiently identified and removed, reducing the impact of muon-induced backgrounds while minimizing the loss of exposure.

It is important to note that the muon rate decreases with the detector volume. The estimate adopted in this work is therefore conservative, as it was calculated assuming a detector volume of 1 m$^3$, comparable to that of the CONUS experiment (1.4 m$^3$). However, the technology proposed here would allow for a reduction of the effective detector volume, which would in turn further decrease the expected muon rate. 

\subsection{Intrinsic Detector Background (IDB)}\label{sec:idb}

The two main sources of intrinsic detector background (IDB) of a CIS are thermal generation of electron-hole  (dark current) and the amplifiers inherent electronic noise (readout noise). Dark current has been widely characterized in both CIS and CCD devices and it manifests as the spontaneous appearance of electrons in detector's pixels. This electron generation obeys a poissonian probability distribution given by Eq.~\ref{eq:darkcurrent}. The number of electrons per pixel is $k$ and $\mu_{DC}$ is the expected (mean) number of occurrences during a given time interval. The parameter $\mu_{DC}$ depends on the exposure time of an image via $fps$ by means of $\mu_{DC}=\frac{\lambda_{DC}}{fps}$, where $\lambda_{DC}$ is the expected number of electrons per pixel per unit of time. 

\begin{equation}\label{eq:darkcurrent}
    P_k=\frac{\mu_{DC}^ke^{-\mu_{DC}}}{k!}
\end{equation}

Another source of IDB is light emission from device transistors, effect also known as electronics glowing. These photons can further interact with the silicon producing electron-holes pairs. For the case of CCD devices, the SENSEI experiment has measured this contribution obtaining 0.36$\times$10$^{-4}$e$^{-}$/pix/day for one transistor~\cite{PhysRevApplied.17.014022}. In CIS, electron-holes pairs produced by dark current and glowing are generated simultaneously, being dark current the dominant. Therefore all measurements take into account both contributions and are reported simple as dark current. 
For example, the CIS reported in~\cite{ma20154mp} has a $0.15\,e^-$/pix/s of $\lambda_{DC}$ at $-20^\circ$C. The \mbox{qCMOS} camera from Hamamatsu has reported a dark current of $0.016\,e^-$/pix/s at $-20^\circ$C~\cite{hamamatsu_technote}.

On the other hand, the electronic readout noise follows the gaussian distribution of Eq. \ref{eq:readout}, with a standard deviation given by the readout noise $\sigma_{\scriptstyle \mathrm{RN}}$. 

\begin{equation}\label{eq:readout}
    G(x)=\frac{1}{\sqrt{2\pi}\sigma_{RN}}e^{-\frac{x^2}{2\sigma_{RN}^2}}
\end{equation}

\begin{figure}[ht!]
    \centering
    \includegraphics[width=\columnwidth]{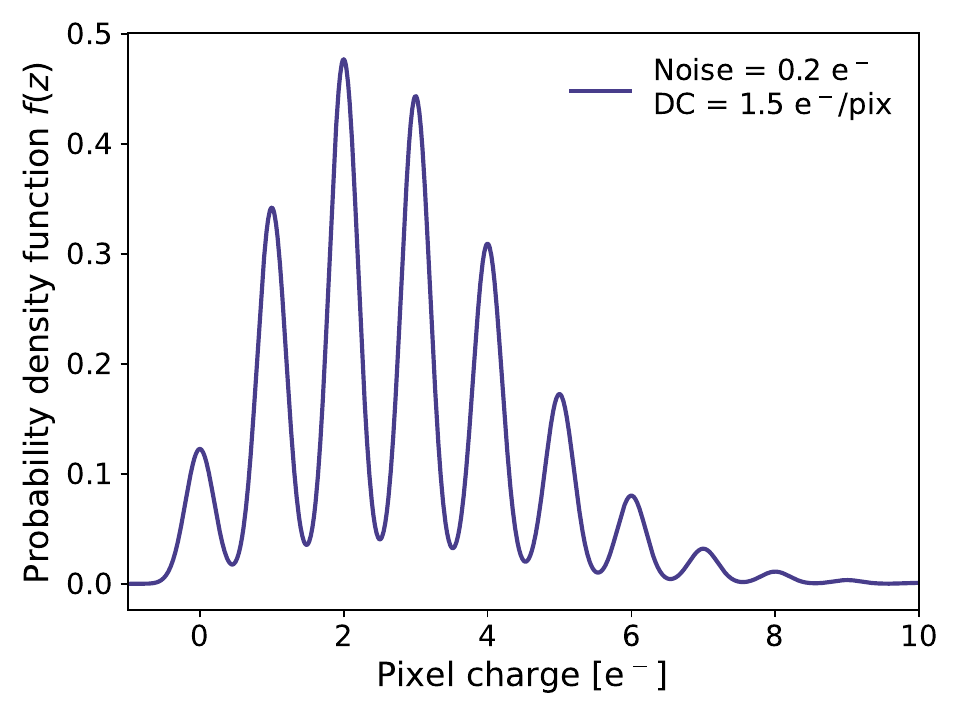}
    \caption{Pixel value PDF from Eq.~\ref{eq:joint_dist} considering only the IDB contributions. Calculated for a detector with a readout noise $\sigma_{RN}$ of 0.2~$e^-$ and dark count $\mu_{DC}$ of 1.5~$e^-$/pix.} 
    \label{fig:clasic_dist}
\end{figure}

The value $z$ of the measurement of a single pixel considering only IDB contribution is then given by the joint probability density function (PDF) of both the readout noise and the dark current. This is given by the convolution of both Eq.~\ref{eq:darkcurrent} and Eq.~\ref{eq:readout} yielding Eq.~\ref{eq:joint_dist}. Fig.~\ref{fig:clasic_dist} displays an example of $f(z)$ for representative detector parameters, where the discrete electron peaks are clearly resolved due to the sub-electron readout noise. 

\begin{equation}\label{eq:joint_dist}
    f(z)=\sum_k\frac{\mu_{DC}^ke^{-\mu_{DC}}}{k!}\frac{1}{\sqrt{2\pi}\sigma_{RN}}e^{\frac{-(z-k)^2}{2\sigma_{RN}^2}}
\end{equation}

\begin{figure}[ht!]
    \centering
    \includegraphics[width=\columnwidth]{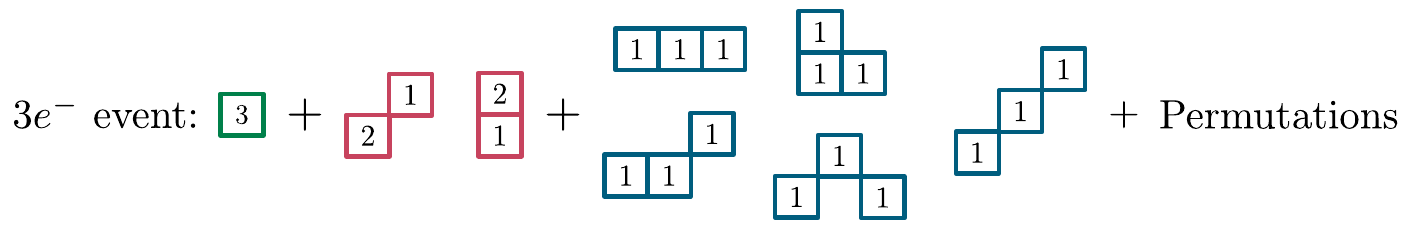}
    \caption{All possible configurations of an event having 3$e^-$.}
    \label{fig:example_poly}
\end{figure}

Electron–hole pairs produced by \cevens's interactions are initiated in a very small volume, determined by the range of the nuclear recoil~\cite{Haro2020}. In depleted silicon, these electrons are drifted toward the collection well of each pixel. During this drift, the electrons also spread laterally due to diffusion. As a result, the charge is collected over a few pixels. For example, in \mbox{250\,$\mu$m-thick}, fully depleted CCDs, the maximum spread expected for bulk events is \mbox{$\sim$6$\,\mu$m}~\cite{Haro2020}. Therefore considering pixels of \mbox{15$\times$15\,$\mu$m$^2$}, about \mbox{99\%} of the signal charge is collected at most in four pixels. Therefore the events in acquired images are defined as groups of connected pixels with their value above a given threshold, chosen as four times the readout noise. 
To properly model the IDB at the event level, it is necessary to extend the analysis beyond the single-pixel probability distribution and account for all possible shapes of connected pixels that can randomly appear in the images by dark current and readout noise. For this purpose, we implement a probability IDB model described in Appendix~\ref{appendix:polykings}. 
This model is based on the counting of geometrical shapes known as \textit{polykings}, that can form randomly in the image. It allows to compute the probability of forming events with arbitrary number of connected pixels. 
An example, all the geometrical shapes of an event with a total of 3~$e^-$ is shown in Fig.~\ref{fig:example_poly}. To get the probability of the event it is sum over all shapes and their permutations. The IDB model has been applied to predict background events from IDB that have a number of pixels in the same range of the signal events, this is, at most four pixels.

\begin{figure*}[ht!]
    \centering
    \includegraphics[width=0.85\textwidth]{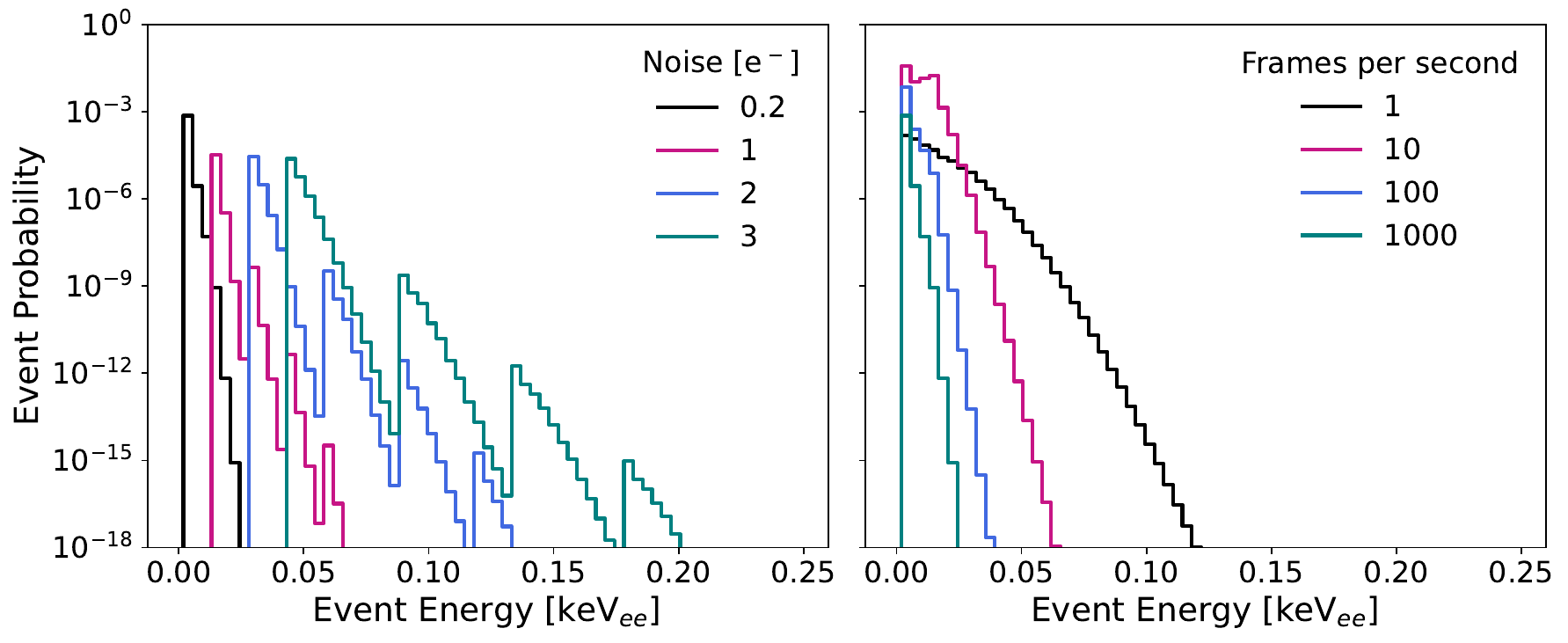}
    \caption{Event probability $P_{\rm IDB}$ computed using the IDB model. Left: sweep in readout noise for a frame rate of 1000\,fps and a dark current of $1.5\,e^-$/pix/s. The threshold is set to four times the readout noise, which produces the saw-like structures. Right: sweep in the number of frames per second with a constant readout noise of $0.2\,e^-$ and a dark current of $1.5\,e^-$/pix/s.}
    \label{fig:noise_dc_sweep}
\end{figure*}

As input parameters the model receives the readout noise $\sigma_{RN}$ and the dark rate $\mu_{DC}$. The result is the probability of getting an event of a given energy of at most four pixels in size. As example cases, Fig.~\ref{fig:noise_dc_sweep} illustrates the resulting probability for two different conditions. The left panel is the case of sweeping the readout noise for a dark rate of 1.5$\times$10$^{-3}$\,$e^-$/pix, that could be the case of a dark current of 1.5\,$e^-$/pix/s at 1000 fps. The right panel present the case for a readout noise of 0.2\,$e^-$ but with different frame rate at a constant dark current.

The expected number background events from IDB ($N_{\rm IDB}$) with a given number of electrons ($n_{e^-}$) is given by Eq.~\ref{eq:Nidb}, it is obtained by multiplying the event probability ($P_{\rm IDB}$) by the live acquisition time ($T$), the number of frames per second ($fps$), and the number of pixels ($N_{\rm pix}$).

\begin{equation} \label{eq:Nidb}
    N_{\rm IDB}(n_{e^-}) = P_{\rm IDB}(n_{e^-},\lambda_{\rm DC},\sigma_{\rm RN})\cdot T \cdot fps \cdot N_{\rm pix}
\end{equation}

\section{\cevens\ Detection with CIS-Based Detector}\label{sec:detection}

\subsection{The \cevens\ Signal in a Silicon Target}

The total reactor antineutrino flux is obtained by summing the contributions from all fissile isotopes, weighted by their respective fission fractions $\alpha_i$, and normalizing to the fission rate. The differential flux at a distance $d$ from the reactor core can be written as~\cite{huber2011determination}
\begin{equation}
    \frac{d\Phi}{dE_{\bar{\nu}_e}} 
    = \frac{n_f}{4\pi d^2}
    \left(
    \sum_{i} \alpha_i 
    \frac{dN_{\bar{\nu}_e}^{(i)}}{dE_{\bar{\nu}_e}}
    \right),
    \label{eq:reactor_flux}
\end{equation}
where $n_f$ is the number of fissions per second given by Eq. \ref{eq:nf}, $W_{\rm th}$ is the reactor thermal power and $\langle E_{\rm fis} \rangle$ is the average energy released per fission. 

\begin{equation}\label{eq:nf}
    n_f = \frac{W_{\rm th}}{\langle E_{\rm fis} \rangle},
\end{equation}

Tabulated spectral data has been used for antineutrino energies below 2~MeV ~\cite{huber2011determination}. At higher energies, the spectrum is commonly described by the phenomenological parameterization from~\cite{vogel1989neutrino}. 

The quantity $W_{th}/d^2$ can be defined as an effective signal strength $H$ (in units of GW$_{\rm th}$/m$^2$, with $d$ in meters)  and factored out from the antineutrino flux, the CEvNS event rate will scale linearly with this quantity $H$. This definition has been adopted in this work to understand different experimental configurations. 
The CEvNS event rate is obtained by convolving the antineutrino flux with the differential cross section and integrating over the kinematically allowed neutrino energies. The differential recoil rate as a function of nuclear recoil energy $E_R$ is therefore

\begin{equation}\label{Eq:Ev_rate}
\frac{dR}{dE_R}(E_R)=
\int_{E_{\bar{\nu}_e}^{\min}}^{\infty}
 \frac{d\Phi}{dE_{\bar{\nu}_e}}
\frac{d\sigma}{dE_R}(E_{\bar{\nu}_e},E_R)
dE_{\bar{\nu}_e},
\end{equation}

The lower integration limit corresponds to the minimum neutrino energy required to produce a recoil of energy $E_R$,

\begin{equation}
E_{\bar{\nu}_e}^{\min}=
\frac{E_R+\sqrt{E_R^2+2 m E_R}}{2},
\end{equation}

\noindent with $m$ being the nuclear mass in MeV, in our case the silicon nucleus's mass.

For CEvNS, the differential cross section in the Standard Model can be written in a simplified form as

\begin{equation}
\frac{d\sigma}{dE_R}=
\frac{G_F^2}{4\pi}
Q_W^2
F^2(q^2)
m
\left(
1-\frac{m E_R}{2E_{\bar{\nu}_e}^2}
\right),
\end{equation}

\noindent where $G_F$ is the Fermi constant, 

\begin{equation}
Q_W = N - (1 - 4 \sin^2\theta_W) Z
\end{equation}
is the weak nuclear charge, and $F(q^2)$ is the nuclear form factor accounting for finite-size effects at momentum transfer $q$. In the case of silicon $Z=14$ and $N=14$. The variables $E_{\bar{\nu}_e}$ and $E_R$ denote the incoming neutrino energy and nuclear recoil energy, respectively.

The signal rate from the detector is calculated by including the quenching factor effect and the resolution including the Fano noise following appendix~\ref{appendix:quench}, this yields the differential rate of events with the energy measured by the detector $dR/dE$. The expected number of signal events $N_{\rm sig}$ is obtained by integrating the differential rate over a selected energy range $[E_{\rm min}, E_{\rm max}]$, and multiplying by the total exposure defined as the product of the detector mass $M$ and the live time $T$. Explicitly, the number of signal events is given by Eq.~ \ref{eq:Nsig}, where $dR/dE$ is expressed in units of \mbox{keV$^{-1}$\,kg$^{-1}$\,day$^{-1}$}, commonly referred to as DRU (differential rate unit). 

\begin{equation}
N_{\rm sig} = M\,T \cdot \int_{E_{\rm min}}^{E_{\rm max}} \frac{dR}{dE}\, dE
\label{eq:Nsig}
\end{equation}
Considering a flat physical background spectrum, the expected number of background events $N_\mathrm{bkg}$ is given by Eq.~\ref{eq:Nbkg}, where $R_\mathrm{bkg}$ is the background rate. 
\begin{equation}
N_\mathrm{bkg} = M\,T \cdot R_\mathrm{bkg} \cdot (E_{\rm max} - E_{\rm min})
\label{eq:Nbkg}
\end{equation}

\begin{figure}[ht!]
    \centering
    \includegraphics[width=\columnwidth]{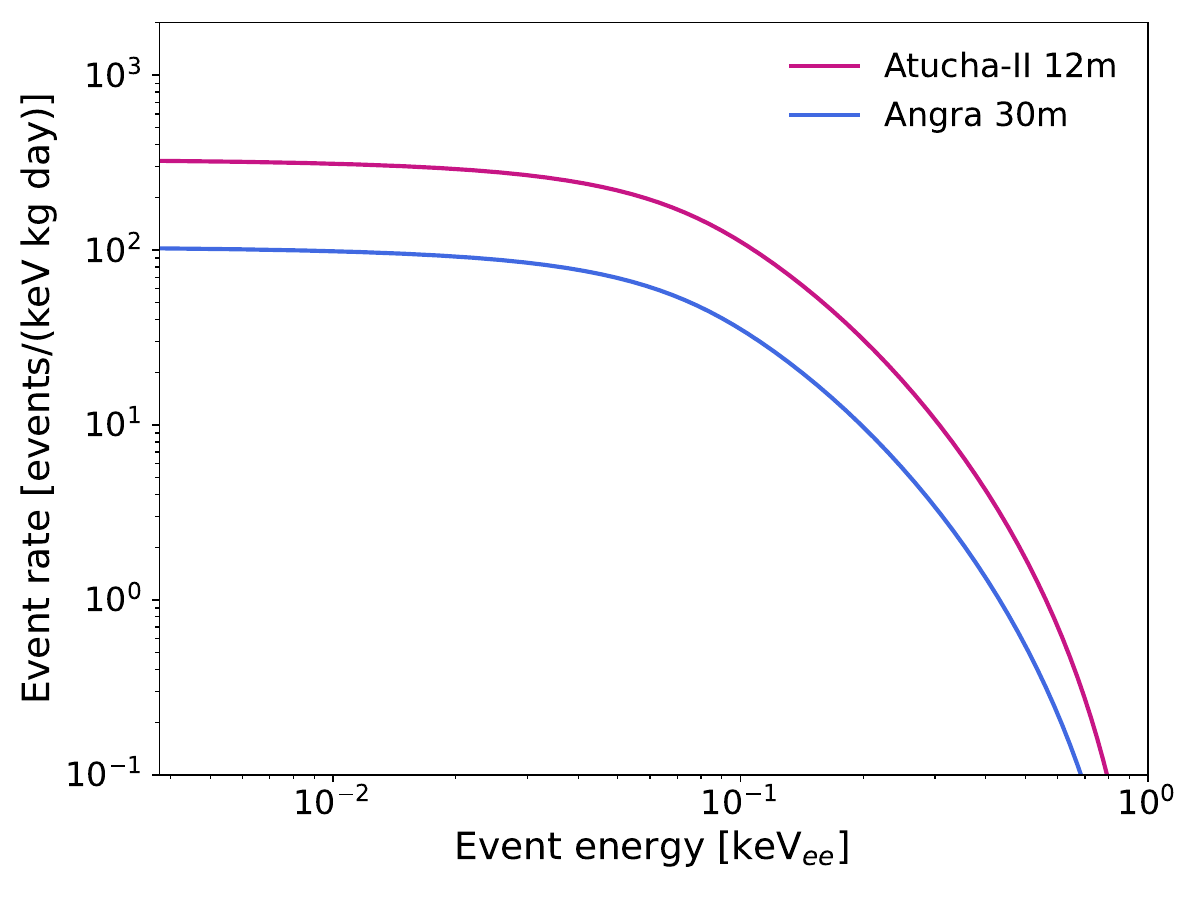}
    \caption{Expected signal rate for two experimental configuration. The effective signal strength are $H_1=0.014$ and $H_2=0.004$.}
    \label{fig:Ev_rate_expected_EM}
\end{figure}

Two representative experimental configurations are considered both characterized by the reactor thermal power and the detector–core distance. The first benchmark, $H_1 = 0.014$~GW$_{\rm th}$/m$^2$, corresponds to near-core, high-flux setups such as ATUCHA~\cite{Depaoli2024b}, CONUS~\cite{ackermann2025direct}, NUCLEUS~\cite{goupy2023exploring}, and PROSPECT~\cite{ashenfelter2019prospect} which achieve comparable effective flux factors. The second, $H_2 = 0.004$~GW$_{\rm th}$/m$^2$, represents configurations with reduced neutrino flux, either due to larger baseline, lower reactor power, or both. This regime is similar to that of CONNIE, NEON, and TEXONO. The expected signal rate $dR/dE$ for each configuration is shown in Fig.~\ref{fig:Ev_rate_expected_EM}.

\begin{figure}[ht!]
    \centering
    \includegraphics[width=\columnwidth]{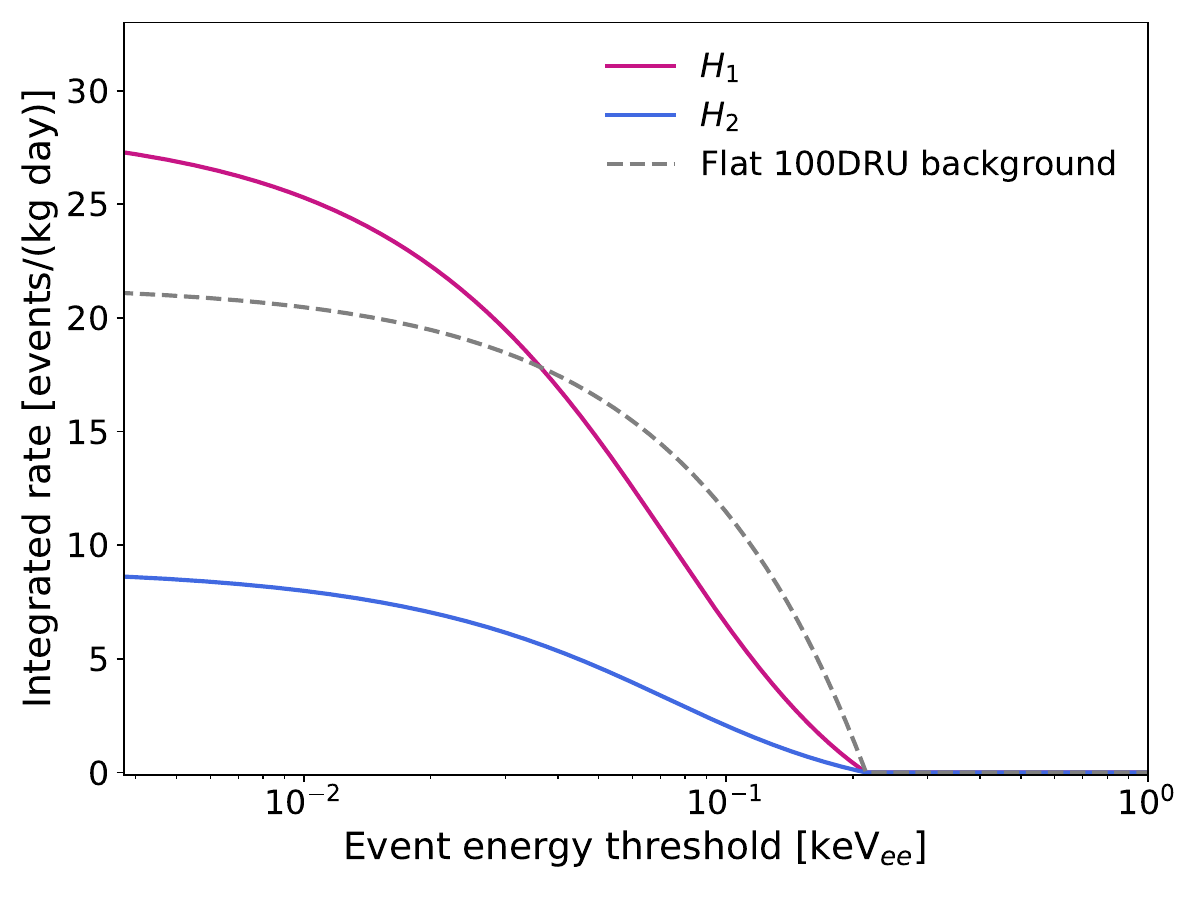}
    \caption{Expected number of events versus detector energy threshold $E_{\min}$. Two experimental cases are shown, $H_1$ and $H_2$. As comparison the integration of a flat background of 100~DRU is included using a maximum integration limit $E_{\max}$ of 0.215~keV.}
    \label{fig:Ev_rate_w_bkg}
\end{figure}

Fig.~\ref{fig:Ev_rate_w_bkg} shows the results for both experimental scenarios $H_1$ and $H_2$ compared to a flat physical background of 100~DRU. The upper integration limit $E_{max}$ has been set at the value to 0.215~keV, above which none \cevens\ events are expected. In the case of $H_1$, with an energy threshold below 0.035~keV, more signal events are expected against the flat background making it an interesting prospect for a fast observation of the \cevens\ signal.

\subsection{Expected Significance in a Nuclear Reactor Site}

To assess the discovery potential of a CIS-based experiment, we treat the measurement as a single-bin counting experiment in the energy range $[E_{\rm min}, E_{\rm max}]$. The expected number of signal events $s$ is given by Eq.~\ref{eq:Nsig}, while the expected background $b$ is the sum of the physical background of Eq.~\ref{eq:Nbkg} and the IDB contribution of Eq.~\ref{eq:Nidb}, integrated over the same window. 

Since the expected event counts are small, the Gaussian approximation $S = s/\sqrt{b}$ is not adequate. Instead, we adopt the median discovery significance $Z$ for a Poisson counting experiment with known background, obtained from the profile-likelihood ratio in the asymptotic (Asimov) approximation~\cite{Cowan:2010js},
\begin{equation}\label{eq:significance}
    Z = \sqrt{\,2\left[\,(s+b)\,
    \ln\!\left(1+\frac{s}{b}\right) - s\,\right]}\,,
\end{equation}
which remains accurate even for a few expected events and reduces to $s/\sqrt{b}$ in the limit $s \ll b$.

Since both $s$ and $b$ scale linearly with the exposure $M\,T$, Eq.~\ref{eq:significance} scales as $\sqrt{M\,T}$, so the exposure required to reach a given significance can be obtained directly for any experimental scenario. The dependence of the expected number of signal events on the detector energy threshold $E_{\rm min}$ follows from Eq.~\ref{eq:Nsig}. 

\subsubsection{Optimal spectrum energy range}

\begin{figure}[ht!]
    \centering
    \includegraphics[width=\columnwidth]{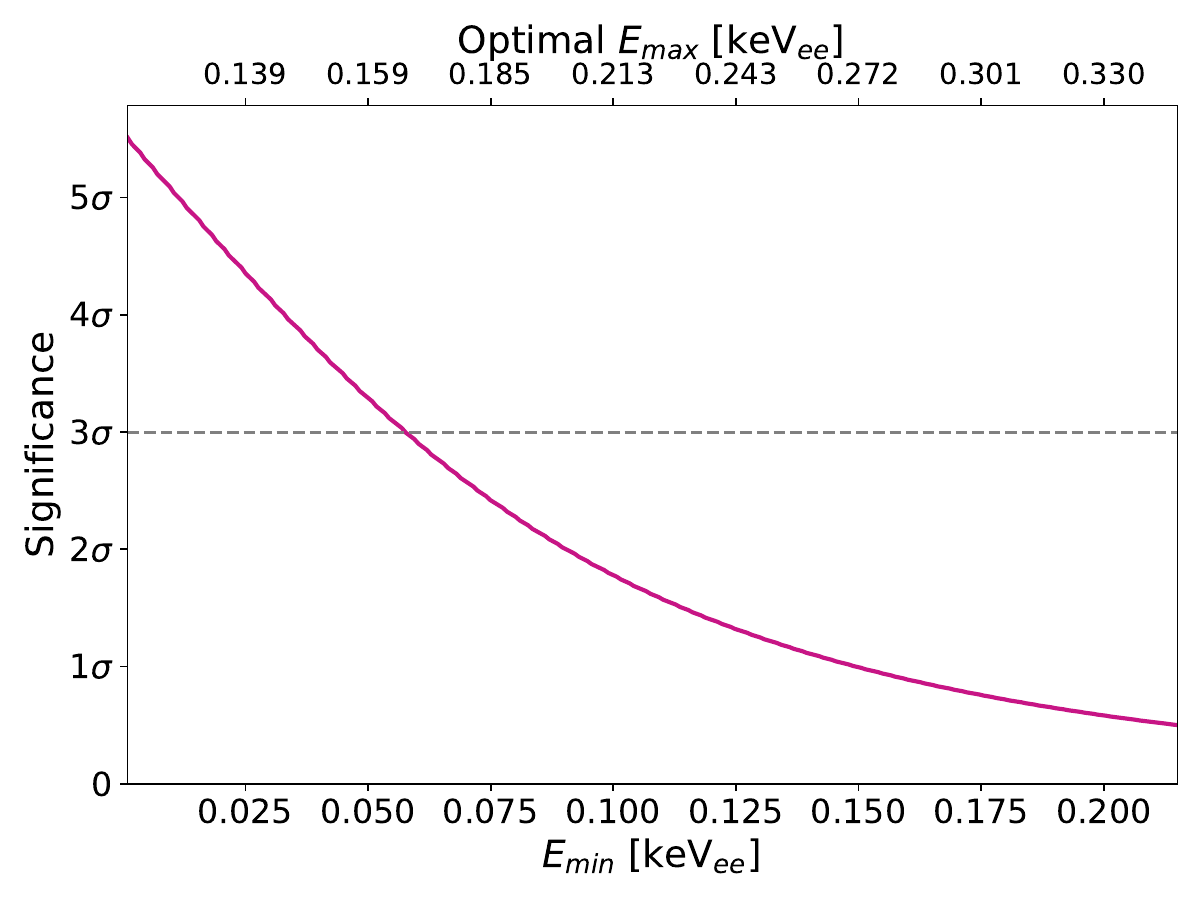}
    \caption{Maximum significance as a function of the optimal spectrum range $[E_{\min}, E_{\max}]$ for the experimental condition $H_1$. A flat background of \mbox{100~DRU} is assumed. The significance is normalized to $\rm1~kg \cdot day$ and it scales with the square root of the exposure ($\sqrt{M\,T}$). A significance of $3\sigma$ is achieved for the window $\rm [0.0576,0.167]\,keVee$.}
    \label{fig:optimal_significanse}
\end{figure}

\begin{figure}[ht!]
    \centering
    \includegraphics[width=\columnwidth]{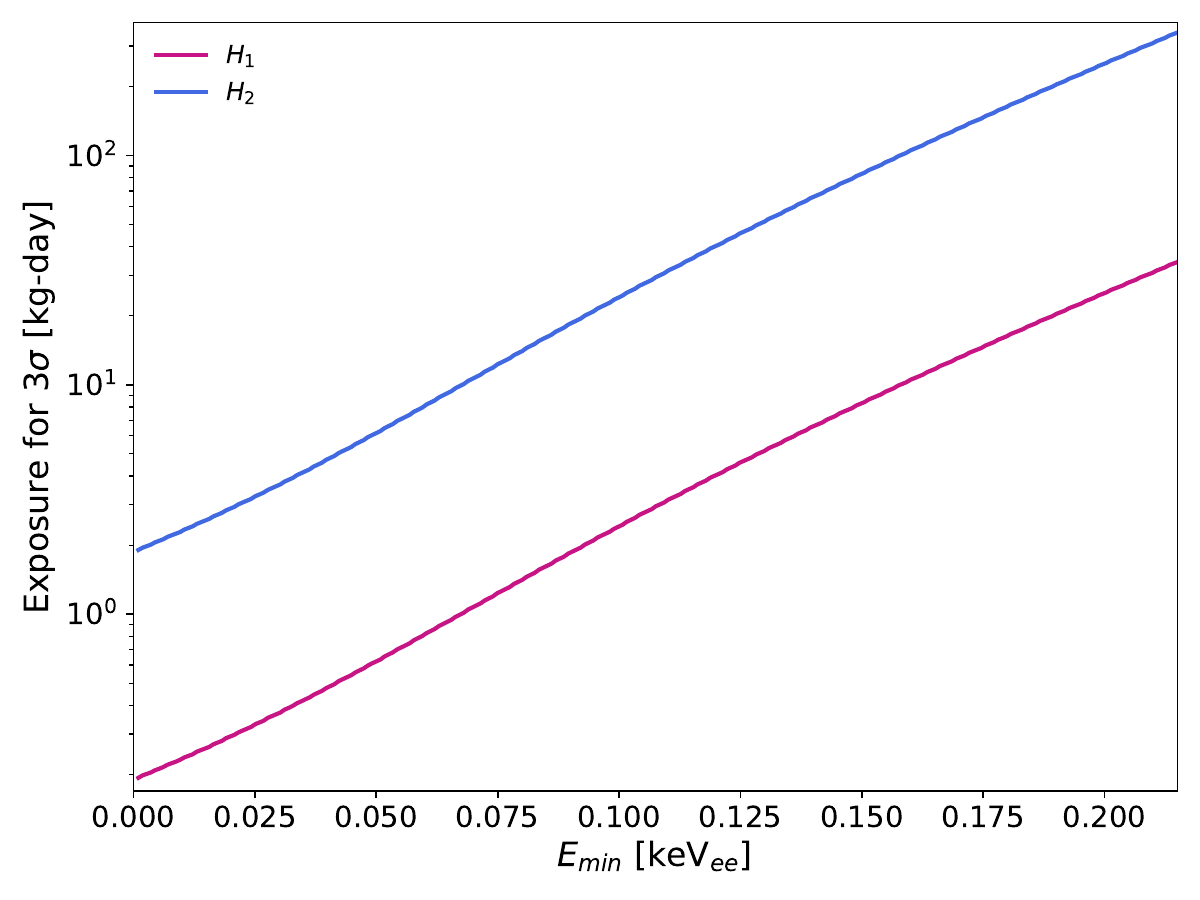}
    \caption{Exposure required to reach a $3\sigma$ significance as a function of the detector energy threshold $E_{\min}$. The optimal integration range is used. A flat background spectrum of 100~DRU is assumed under $H_1$ and $H_2$ experimental conditions.}
    \label{fig:min_thre_with_optimal}
\end{figure}

In order to determine the spectrum energy range that maximize the significance for the \cevens\ detection, we perform a two-dimensional scan over the lower and upper integration limits of Eq.~\ref{eq:Nsig}, $E_{\min}$ and $E_{\max}$ respectively. For each range $[E_{\min}, E_{\max}]$, we compute the expected number of signal and background events, and calculate the significance under the assumption of a flat 100\,DRU background using Eq.~\ref{eq:significance}. The optimal range corresponds to the balance between retaining a large fraction of the \cevens\ spectrum and suppressing the accumulated background that increases linearly with the size of the energy range. The result from the scan for the case of $H_1$ is shown in Fig.~\ref{fig:optimal_significanse}. Given a detector energy threshold $E_{\min}$, Fig.~\ref{fig:optimal_significanse} shows the maximum significance that is achieved for the optimal $E_{\max}$. The significance is for an exposure of one $\rm kg \cdot day$, and it scales with the square root of the exposure ($\sqrt{M\,T}$). In Fig.~\ref{fig:min_thre_with_optimal} the exposure required to reach a significance of $3\sigma$ for each detector energy threshold is shown for both experimental conditions $H_1$ and $H_2$.

\section{Prospects of a Feasible CIS-Based Detector}\label{sec:viable}

\begin{figure}[ht!]
    \centering
    \includegraphics[width=\columnwidth]{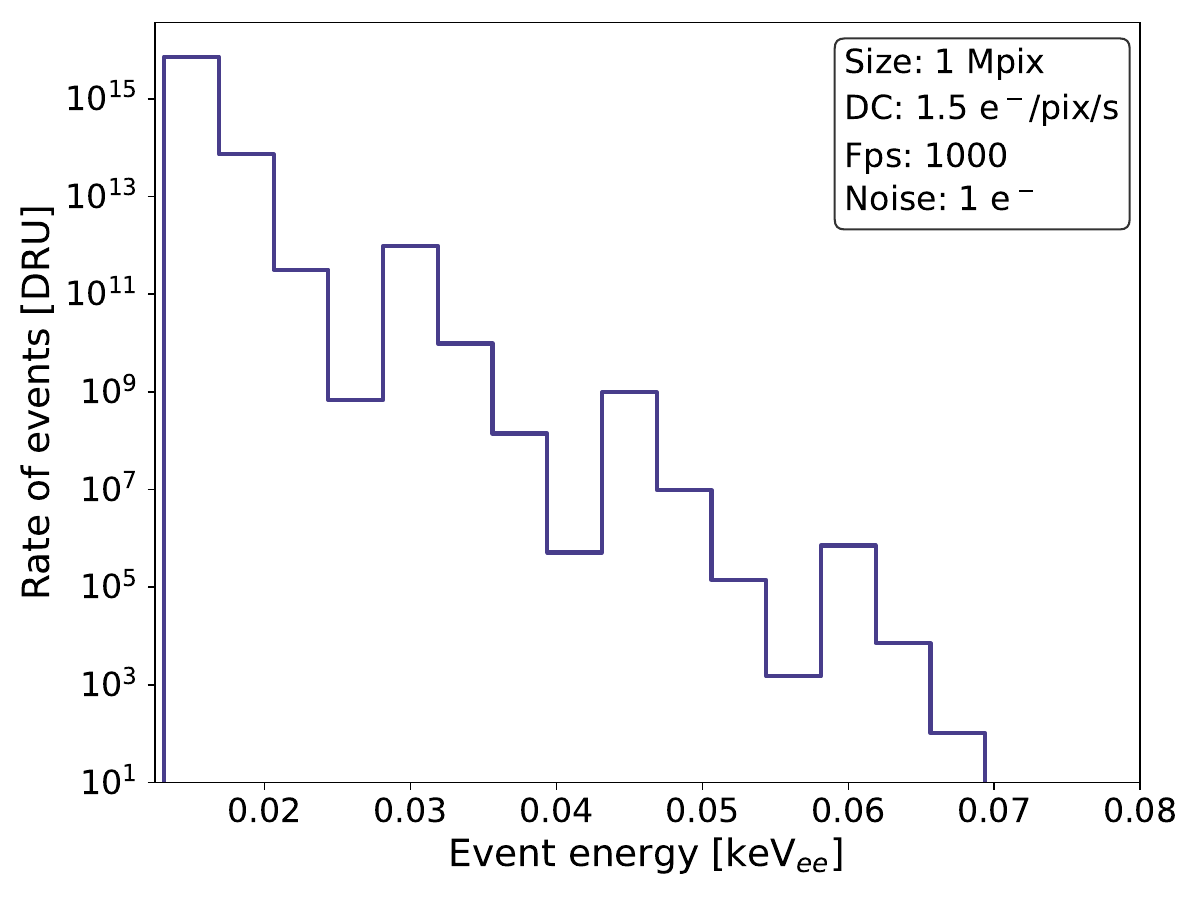}
    \caption{Event rate obtained with the IDB model for the detector proposed in table \ref{tab:detector} (1~Mpix detector, dark current of 1.5~$e^-$/pix/s, 1000~fps, and 1~$e^-$ readout noise). The distribution is shown in DRU units as a function of event energy.}
    \label{fig:sensor_particular}
\end{figure}

Having modeled the \cevens\ signal and quantified the significance achieved in the presence of a \mbox{100\,DRU} physical background, this section considers the IDB of a CIS-based detector with the characteristics listed in Table~\ref{tab:detector}. Based on the review of modern CIS technologies presented in Section~\ref{sec:cis}, we consider the detector specified in Table~\ref{tab:detector} to be technologically feasible. The assumed noise and dark current levels are within the range reported for existing pixels, and the noise could be further reduced through the use of non-destructive charge readout. Achieving the target frame rate relies on hybrid integration with fully parallel pixel readout. High density, low background packing has been already developed for CCD detectors and could be re-engineered for CIS. Packing 256, 1\,Mpix CCDs in a compact volume of \mbox{$\rm150\times150\times50\,mm^3$} has been demonstrated in OSCURA experiment~\cite{botti2025multichannel,aguilar2022oscura}. Fig.~\ref{fig:sensor_particular} shows the resulting IDB obtained with the model described in section \ref{sec:idb}. 

\begin{table}[]
\begin{tabular}{l l}
\hline \hline
Readout noise            & 1\,$e^-$          \\ 
Dark current             & 1.5\,$e^-$/pix/s  \\ 
Silicon thickness        & 725\,$\mu$m    \\ 
Number of pixels         & 1\,Mpix        \\ 
Pixel size               & $\rm 15\times15\,\mu m^2$ \\
Frames per second        & 1000           \\ 
Number of CISs           & 256             \\ 
Total silicon mass       & 97\,g          \\ \hline \hline
\end{tabular}
\caption{Key requirements for a feasible and compact CIS-based neutrino detector.}
\label{tab:detector}
\end{table}

\begin{figure}[ht!]
    \centering
    \includegraphics[width=\columnwidth]{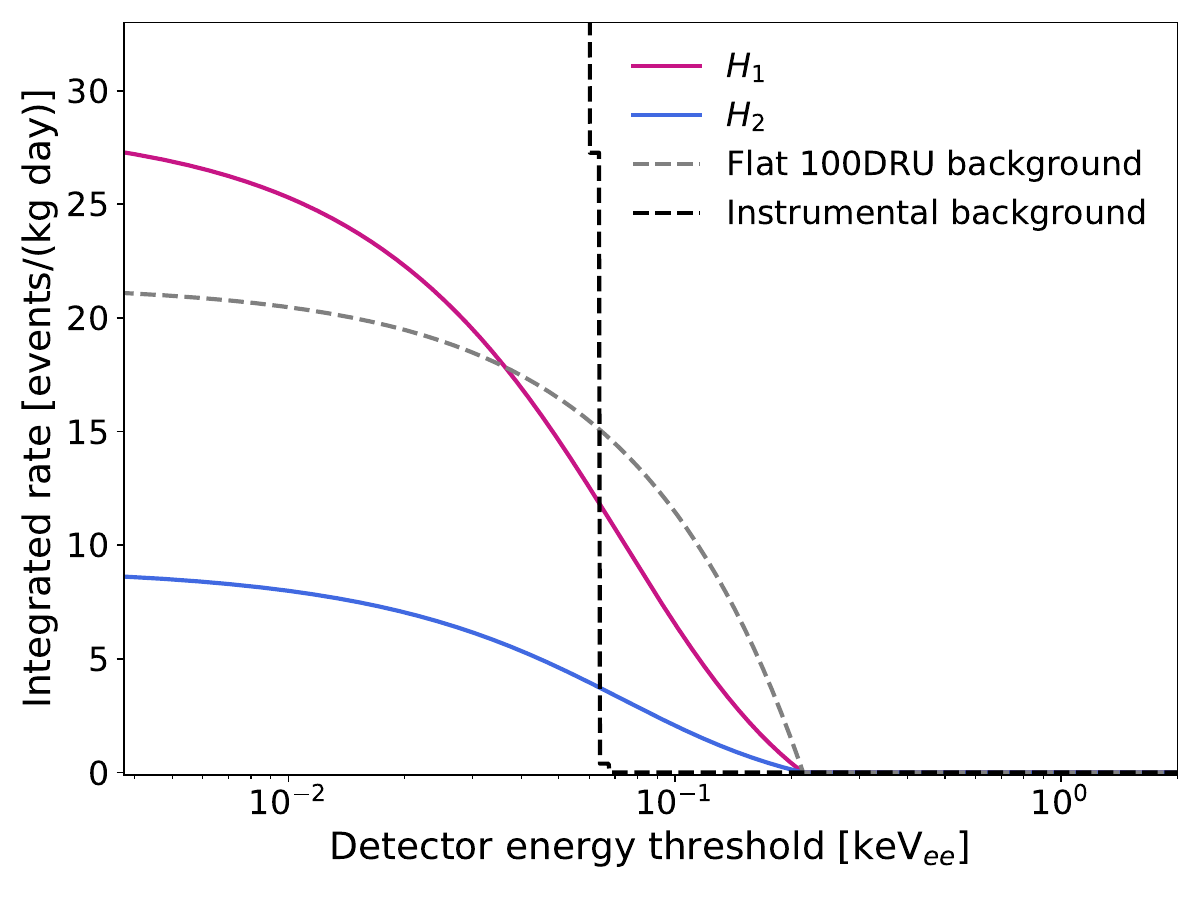}
    \caption{Integrated event rate as a function of the detector energy threshold. The solid curves correspond to the expected signal for the two benchmark flux scenarios $H_1$ and $H_2$, while the dashed gray curve represents a flat 100~DRU background. As can be observed detector energy threshold must be over 0.064~keV to avoid the contribution from the IDB.}
    \label{fig:integrated_rate_w_instr}
\end{figure}

\begin{figure}[ht!]
    \centering
    \includegraphics[width=\columnwidth]{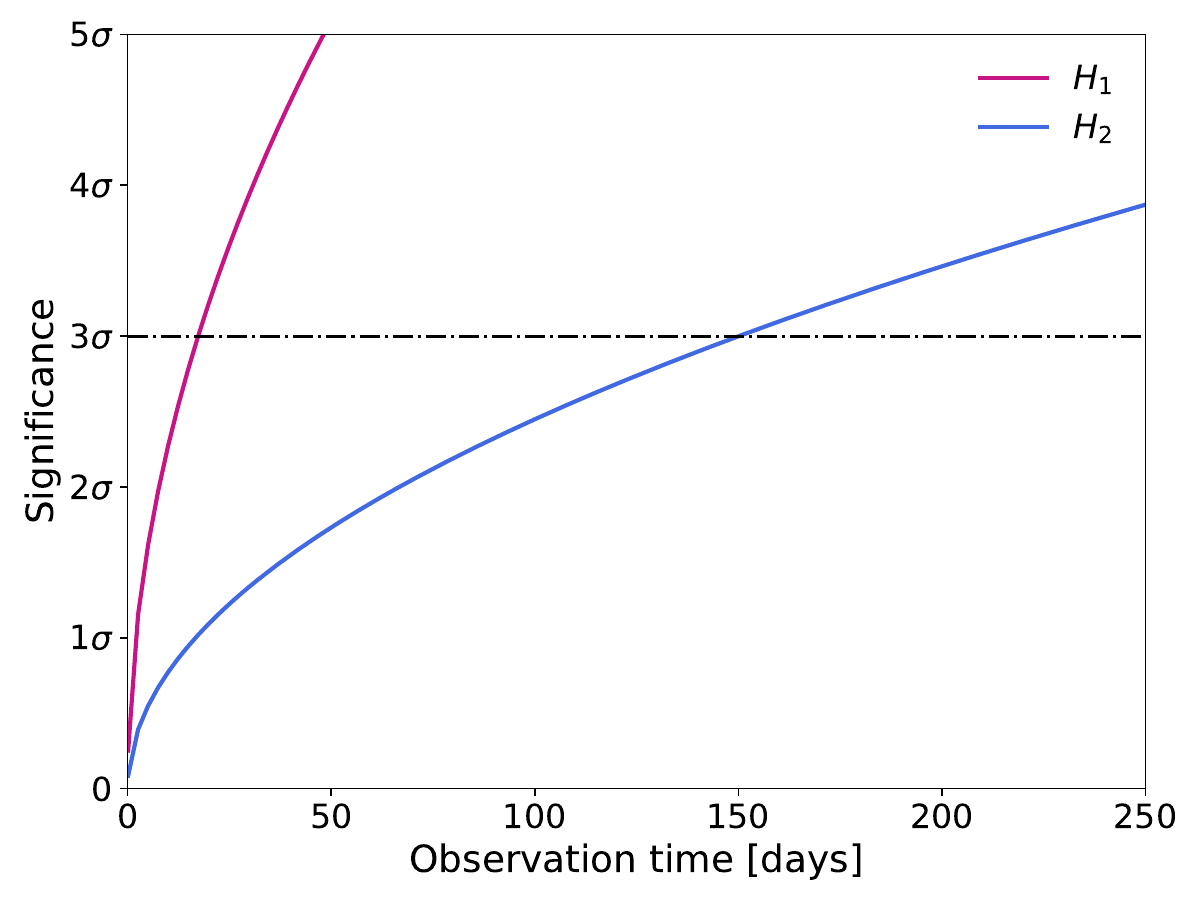}
    \caption{Significance as a function of observation time for the detector presented in table \ref{tab:detector}. The $3\sigma$ significance requirement is achieved in 17~days for the case of $H_1$ and 150~days for the case of $H_2$.}
    \label{fig:expo_2sig_final_instr}
\end{figure}

In Fig.~\ref{fig:integrated_rate_w_instr} is the comparison of the integrated event rate between the \cevens\ signal in scenarios $H_1$ and $H_2$, the flat physical background of 100~DRU, and the IDB contribution. As can be observed, detector threshold must be over \mbox{0.064~keV} to supress IDB contribution. In order to estimate the exposure required to reach a 3$\sigma$ of significance, from figure \ref{fig:optimal_significanse} an optimal spectrum range of \mbox{[0.065, 0.158~keV]} is obtained. The results are shown in Fig.~\ref{fig:expo_2sig_final_instr}. For the scenario $H_1$ to reach \cevens\ an observation time of 17~days is necessary. In contrast, for $H_2$ the observation time is equal to 150~days. The inefficiency due to the live time given by the active shielding is already included in the calculation and increases the observation time by 20\%.

\section{Discussion and Conclusion}\label{sec:conclusion}

The presented analytical framework has been applied to analyze other detector configurations. Table \ref{tab:final_result} summarize the results for $H_1$ scenario assuming a physical background of 100~DRU and $\sim$100\,g of detector mass. As it is possible to observe, in all configuration $3\sigma$ of significance is reached in less than 50~days. In case of $H_2$, about an order of magnitude larger than $H_1$ due to its lower signal expectation. This further demonstrates that reactor \cevens\ detection is already feasible with 1~$e^-$ noise CIS in a short time scale, even while progress towards sub-electron performance would substantially reduce the time to the \cevens\ detection.

\begin{table*}[ht!]
\centering
\begin{tabular}{|c|c|c|c|c|c|}
\hline
\hline
Readout noise [$e^-$] & \shortstack{Dark current [$e^-/\text{pix}/s$]} & \shortstack{Frames per\\ second} 
&  \shortstack{Days to reach \\ $3\sigma$ (100\,g)} 
& \shortstack{Signal \\ events} 
& \shortstack{Detector energy \\ threshold [keVee]} \\
\hline
\hline
\multirow[c]{6}{*}{1}
& \multirow[c]{3}{*}{1.5}
& 200  & 44 & 38 & 0.072 \\
\cline{3-6}
&      & 500  & 21 & 19 & 0.071  \\
\cline{3-6}
&      & 1000 & 17 & 15 & 0.072 \\
\cline{2-6}

& \multirow[c]{3}{*}{$1.5\times10^{-3}$}
& 200  & 35 & 39 & 0.071 \\
\cline{3-6}
&      & 500  & 21 & 19  &  0.071 \\
\cline{3-6}
&      & 1000 & 17 & 15 & 0.071 \\
\hline
\hline

\multirow[c]{6}{*}{0.5}
& \multirow[c]{3}{*}{1.5}
& 200  &  22& 29 & 0.04 \\
\cline{3-6}
&      & 500  & 12 & 17 & 0.037 \\
\cline{3-6}
&      & 1000 & 10 & 14 & 0.037 \\
\cline{2-6}

& \multirow[c]{3}{*}{$1.5\times10^{-3}$}
& 200  & 22 & 32 & 0.034 \\
\cline{3-6}
&      & 500  & 12  & 17 & 0.034 \\
\cline{3-6}
&      & 1000 & 10 & 14 & 0.034 \\
\hline
\hline

\multirow[c]{6}{*}{0.2}
& \multirow[c]{3}{*}{1.5}
& 200  & 16 & 24 & 0.033 \\
\cline{3-6}
&      & 500  & 12 & 18 & 0.029  \\
\cline{3-6}
&      & 1000 & 7 & 11 & 0.027 \\
\cline{2-6}

& \multirow[c]{3}{*}{$1.5\times10^{-3}$}
& 200  & 16 & 30 & 0.017 \\
\cline{3-6}
&      & 500  & 9  & 16 &0.015 \\
\cline{3-6}
&      & 1000 & 7 & 13 & 0.015  \\
\hline
\hline
\end{tabular}
\caption{Time to reach a $3\sigma$ of significance for scenario $H_1$ and different detector parameters: frame rate, dark current and readout noise. The exposure already accounts for the detector live time at each frame rate. In case of $H_2$, the time is roughly an order of magnitude larger than $H_1$ due to its lower signal expectation.}
\label{tab:final_result}
\end{table*}

From Table~\ref{tab:final_result}, two main applications can be identified: one primarily oriented toward technological developments and applied neutrino detection, and another focused on fundamental physics searches. Readout noise of 1~$e^-$ and 200~fps already enables reactor monitoring applications, detector characterization, and the validation of low-energy response models in silicon-based devices~\cite{huber2026}. The second category is oriented toward searches for physics beyond the Standard Model. These studies require enhanced sensitivity to the lowest-energy ionization signals, which in turn demands sub-electron readout noise. Achieving this regime would significantly improve the signal-to-background ratio and extend the sensitivity of searches for non-standard neutrino interactions (like the neutrino magnetic moment), light mediators, or other exotic scenarios like millicharged particles~\cite{mCP_CONNIE_Atucha}. From the previous analysis we consider that CIS technology provides a realistic pathway toward this goal. Additional multiple non-destructive sampling capability allows the readout noise to be reduced below the single-electron level maintaining a high frame rate. Such advances would open the door to precision \cevens\ measurements and competitive probes of new physics in the low-momentum-transfer regime.

\appendix
\section{Analytical Model for Estimating Event Rates in Pixelated Detectors}
\label{appendix:polykings}
In this appendix we present an analytical model for the estimation of event rates in pixelated detectors. With this model an estimate of rate of instrumental events for detectors with different parameters was computed though this work. The model was formulated going from single pixel probabilities to probabilities of clusters with a fixed energy. The probability of having an electron above a certain threshold is given by the integral of Eq.~\ref{eq:joint_dist} from said threshold to infinity. The probability of reading a discrete number of electrons $n$ in a pixel is given by

\begin{equation}\label{eq:prob_bin}
    P(z=ne^-)=P((n-0.5)e^-<z<(n+0.5)e^-)
\end{equation}
which is obtained by integrating Eq.~\ref{eq:joint_dist} in the desired region.

The probability given by Eq.~\ref{eq:prob_bin} considers a single pixel having a certain number of charge due to intrinsic sources (dark current and readout noise). If we now want to think in terms of events i.e. clusters of connected pixels with a given charge, we must add to this simple model information about the image lattice. A single pixel event will be thus defined as a pixel with a given charge and empty neighbors. In a square lattice like the ones from CCD and CMOS image arrays, considering all directions, this corresponds to

\begin{equation}\label{size_1}
    n_1=pq^8,
\end{equation}
here $p$ is the probability for any number of electrons and $q=1-p$. In order to generalize this to events of any size we can write

\begin{equation}\label{eq:size_dist}
    n_s=\sum_{pattern}N_{s,v(s)}p^sq^{v(s)}
\end{equation}
where the information about the cluster geometry is encoded in $N_{s,v(s)}$ which is the numbers of ways a cluster can be rotated or reflected in an image and still be the same cluster, and $v(s)$ which is the number of empty neighbors that depends on the shape of the cluster. The probability of a cluster with a given size will be given by a polynomial in the activation probability $p$ that depends on the noise and dark current of the device. The structures on a square lattice considering neighbors in all directions are a known mathematical entity~\cite{golomb1996polyominoes} called \textit{pseudo-polyominos} or \textit{polykings} (referencing the fact that a king in chess can move in all directions in the board). The shapes that appear in the known Tetris game are a special case of \textit{polykings} called \textit{tetrominos}. To obtain the probability for a given cluster size, we sum over what is known as a \textit{free} Polyking, i.e., over all distinct Polyking configurations that are not related to each other by rigid transformations (translations, rotations, reflections, or glide reflections), meaning that configurations that can be mapped onto one another by picking up and flipping the piece are counted only once. Since the number of possible polykings increases rapidly with the cluster size we will restrict this analysis to events with size four. Coincidentally this is the maximum expected size for the physics events that we are interested in. As a reference for size 4 polykings (tetrakings) there are 22 free patterns, for size 5 (pentakings) there are 94 and 524 for size 6 (hexakings). There is no known closed formula to calculate the number of polykings for each size $s$ and their permutations or $v(s)$ function. 

\begin{figure}[ht!]
    \centering
    \includegraphics[width=\columnwidth]{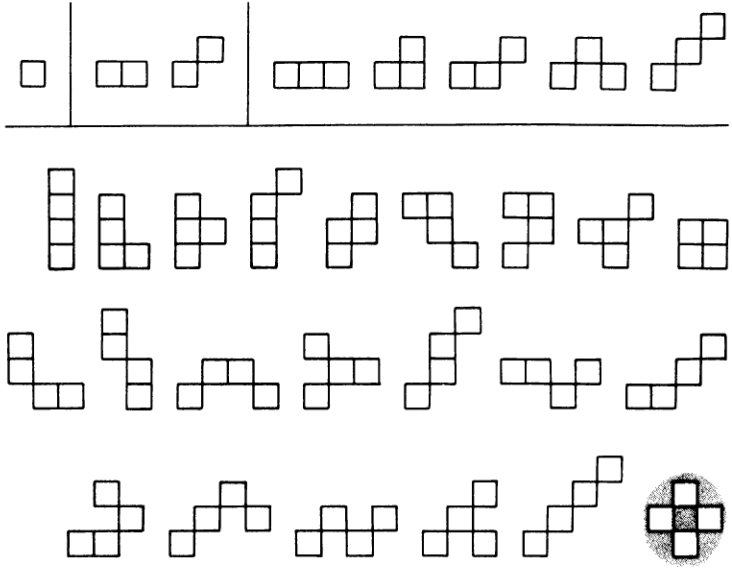}
    \caption{\textit{Free} polykings for sizes 1, 2, 3 and 4 considered for the instrumental background in this work. Figure taken from~\cite{golomb1996polyominoes}.}
    \label{fig:polykings}
\end{figure}

Fig.~\ref{fig:polykings} shows the free polykings that where used in this work up to size 4. The polynomial function given by Eq.~\ref{eq:size_dist} is calculated by doing the permutations for each possible pattern yielding,

\begin{equation}\label{size_2}
    n_2=2p^2(q^{10}+q^{12})
\end{equation}

\begin{equation}\label{size_3}
    n_3=p^3(4q^{15}+2q^{16}+6q^{12}+8q^{14})
\end{equation}

\begin{equation}\label{size_4}
n_4=p^4(21q^{17}+24q^{18}+8q^{19}+q^{12}+28q^{16}+8q^{15}+18q^{14}+2q^{20})
\end{equation}

While Eqs.~\ref{size_1},~\ref{size_2},~\ref{size_3} and~\ref{size_4} provide the probability for a cluster of fixed size $s$, the physically relevant observable is the total collected charge $Q$, expressed in number of electrons. Since different cluster geometries and sizes can correspond to the same total charge, the probability for observing a charge $Q$ is obtained by summing over all cluster configurations that yield that charge.

To compute the probability of observing a charge $Q$, one must sum over all possible realizations of that charge within connected clusters of pixels. For a cluster of fixed size $s$, the total charge is
\begin{equation}
Q = \sum_{i=1}^{s} k_i ,
\end{equation}
where $k_i$ denotes the number of electrons collected in pixel $i$. 

For a given $Q$, this requires considering all integer partitions of $Q$ into $s$ strictly positive integers. As an illustrative example, for $Q=20$ and $s=4$, one must include all partitions of 20 into four positive integers, such as
\[
(5,5,5,5), \quad (5,5,6,4), \quad (7,5,4,4), \quad \ldots
\]
Each partition corresponds to a particular distribution of charge among the pixels of the cluster.

Since the pixels within a cluster are distinguishable by their spatial position, different permutations of a given partition represent distinct configurations and must be counted separately. For instance, for $Q=4$ and $s=3$, the assignments
\[
(2,1,1), \quad (1,2,1), \quad (1,1,2)
\]
correspond to different realizations of the same partition and contribute independently to the total probability. 

Although the number of configurations increases rapidly with $Q$, the sum remains finite for fixed total charge or converges rapidly. The probability of observing charge $Q$ is therefore obtained by summing over
\begin{enumerate}
    \item all cluster sizes $1 \le s \le 4$
    \item all cluster geometries of size $s$
    \item  all charge partitions of $Q$ into $s$ positive integers, including their permutations across pixels.
\end{enumerate}

\section{Quenching factor and detector resolution}\label{appendix:quench}

To relate the differential nuclear recoil rate from Eq.~\ref{Eq:Ev_rate} to the differential ionization rate measured by an APS detector, the nuclear quenching factor must be taken into account. The quenching factor is defined as
\begin{equation}
    Q = \frac{E_I}{E_R},
\end{equation}
where $E_R$ is the nuclear recoil energy and $E_I$ is the ionization energy deposited in the detector. 

At low energies in silicon, the quenching factor has been parameterized as~\cite{chavarria2016measurement}
\begin{equation}
    Q(E_I)=
    \frac{p_3E_I+p_4E_I^2+E_I^3}
         {p_0+p_1E_I+p_2E_I^2},
\end{equation}
with coefficients $p_0=56\,{\rm keV}^3$, $p_1=1096\,{\rm keV}^2$, $p_2=382\,{\rm keV}$, $p_3=168\,{\rm keV}^2$, and $p_4=155\,{\rm keV}$. The differential ionization spectrum is then obtained through the change of variables
\begin{equation}
    \frac{dR}{dE_I}
    =
    \frac{dR}{dE_R}
    \frac{1}{Q}
    \left(
    1-\frac{E_I}{Q}\frac{dQ}{dE_I}
    \right).
\end{equation}

Finally, to write the expected number of events in terms of what is actually measured, assuming that for a given ionization energy $E_I$ the measured energy follows a Gaussian distribution due to detector resolution effects, the observable differential rate as a function of the measured energy $E$ is given by
\begin{equation} \label{eq:sigrate}
    \frac{dR}{dE}
    =
    \frac{
    \int_0^{\infty}
    G(E,E_I;\sigma_I^2)
    \frac{dR}{dE_I}\, dE_I
    }
    {
    \int_0^{\infty}
    G(E,E_I;\sigma_I^2)\, dE_I
    }
\end{equation}
where
\begin{equation}
    G(E,E_I;\sigma_I^2)
    =
    \frac{1}{\sqrt{2\pi\sigma_I^2}}
    \exp\left(
    -\frac{(E-E_I)^2}{2\sigma_I^2}
    \right)
\end{equation}
is the Gaussian response function of the detector.

Here, $E_I$ denotes the true ionization energy and the variance of the energy resolution is modeled as
\begin{equation}
    \sigma_I^2
    =
   s  \, \sigma_{\scriptstyle \mathrm{RN}}^2 
    +
    F\, E_{eh}\, E_I,   
\end{equation}
where $s$ is the size of the event and $\sigma_{\scriptstyle \mathrm{RN}}$  was taken as 2~e$^{-}$ as a conservative value, $F$ is the Fano factor for silicon~\cite{rodrigues2023unraveling}, and $E_{eh}=3.75$~eV is the mean energy required to produce an electron–hole pair in silicon. This step accounts for the intrinsic statistical fluctuations in charge production.

\bibliography{biblio}

\end{document}